\documentclass[aps,superscriptaddress,twocolumn,pra]{revtex4-1}%
\usepackage{amsfonts}
\usepackage{amsmath}
\usepackage{amssymb}
\usepackage{graphicx}
\usepackage{xcolor}
\usepackage{verbatim}
\usepackage{bbm}

\usepackage{hyperref}
\newtheorem{theorem}{Theorem}

\newtheorem{lemma}[theorem]{Lemma}

\newtheorem{proposition}[theorem]{Proposition}

\newcommand{\etal}{\emph{et al.{}}}
\newcommand{\rent}[2]{D\left(#1||#2\right)}

\newcommand{\ket}[1]{|#1\rangle}
\newcommand{\bra}[1]{\langle #1|}
\newcommand{\bracket}[2]{\langle #1|#2\rangle}
\newcommand{\ketbra}[1]{|#1\rangle\langle #1|}
\def\tr{{\rm Tr}}
\def\ot{\otimes}
\def\ep{\epsilon}
\def\eps{\ep}

\usepackage{xspace}
\def\sm{Appendix\xspace}

\usepackage{multibib}

\def\h2{h}

\newcommand{\be}{\begin{eqnarray} \begin{aligned}}
\newcommand{\ee}{\end{aligned} \end{eqnarray} }
\def\ot{\otimes}

\newcommand{\proj}[1]{| #1 \rangle   \langle #1 |}
\newcommand{\mean}[1]{\langle #1 \rangle}
\def\tr{{\rm Tr}}
\def\c{c_{t's'}^{kg}}

\begin{document}
\title{The Resource Theory of Quantum States Out of Thermal Equilibrium}
\author{Fernando G.S.L.\ Brand\~{a}o}
\affiliation{Departamento de F\'isica, Universidade Federal de Minas Gerais, Belo Horizonte, Brazil}
\affiliation{Centre for Quantum Technologies, National University of Singapore, Singapore}
\author{Micha\l~Horodecki}
\affiliation{Institute for Theoretical Physics and Astrophysics, University of Gda\'nsk,  Gda\'nsk, Poland}
\affiliation{National Quantum Information Centre of Gda\'nsk, Sopot, Poland}
\author{Jonathan~Oppenheim}
\affiliation{Department of Applied Mathematics and Theoretical Physics, University of Cambridge, United Kingdom}
\author{Joseph M. Renes}
\affiliation{Institut f\"ur Angewandte Physik, TU Darmstadt, Darmstadt, Germany}
\affiliation{Institut f\"ur Theoretische Physik, ETH Zurich, Z\"urich, Switzerland}
\author{Robert W. Spekkens}
\affiliation{Perimeter Institute for Theoretical Physics, Waterloo, Canada}

\begin{abstract}
The ideas of thermodynamics have proved fruitful in the setting of quantum information theory, in particular the notion that when the allowed transformations of a system are restricted, certain states of the system become useful \emph{resources} with which one can prepare previously inaccessible states. The theory of entanglement is perhaps the best-known and most well-understood resource theory in this sense.
Here we return to the basic questions of thermodynamics using the formalism of resource theories developed in quantum information theory and show that the free energy of thermodynamics emerges naturally from the resource theory of energy-preserving transformations. Specifically, the free energy quantifies the amount of useful work which can be extracted from asymptotically-many copies of a quantum system when using only reversible energy-preserving transformations and a thermal bath at fixed temperature.  The free energy also quantifies the rate at which resource states can be reversibly interconverted asymptotically, provided that a sublinear amount of coherent
superposition over energy levels is available, a situation analogous to the sublinear amount of classical communication required for entanglement
dilution.
\end{abstract}
\volumeyear{year}
\volumenumber{number}
\issuenumber{number}
\eid{identifier}
\date{\today}
\maketitle

Quantum resource theories are specified by a restriction on the quantum operations (state preparations, measurements, and transformations) that can be implemented by one
or more parties. This singles out a set of states which can be prepared under the restricted operations.  If the parties facing the restriction acquire a quantum state outside the restricted set of states, then they can use this state to implement measurements and transformations that are outside the class
of allowed operations, consuming the state in the process. 
Thus, such states are useful resources.  

A few prominent examples serve to illustrate the idea:
if two or more parties are restricted to communicating classically and implementing local quantum operations, then entangled states become a resource~\cite{horodecki_quantum_2009};
if a party is restricted to quantum operations that have a particular symmetry, then states that break this symmetry become a resource~\cite{janzing_quasi-order_2003,gour_resource_2008,marvian_pure_2011};
if a party is restricted to preparing states that are completely mixed and performing unitary
operations, then any state that is not completely mixed, i.e.\ any state that has some purity, becomes a resource~\cite{horodecki_reversible_2003}.  

In this Letter we develop the quantum resource theory of states that are \emph{athermal} (relative to temperature $T$).
 This provides a useful new formulation of nonequilibrium thermodynamics for finite-dimensional quantum systems, and allows us to apply new mathematical tools to the subject. The restricted class of operations which defines our resource theory are those that can be achieved through energy-conserving unitaries and the preparation of any ancillary system in a thermal state at temperature $T$, as first studied by Janzing \etal~\cite{janzing_thermodynamic_2000} in the context of Landauer's principle. Here the ancillary systems can have an arbitrary Hilbert space and an arbitrary Hamiltonian, and may be described as having access to a single heat bath at temperature $T$.   States that are \emph{not} in thermal equilibrium at temperature $T$, that is, which are {athermal}, are the resource in this approach.

Quantum resource theories provide answers to questions such as:
How does one measure the quality of different resource states? Can one particular resource state be
converted to another deterministically? If not, can it be done
nondeterministically, and if so with what probability? What if one has
access to a catalyst?  A particularly fundamental problem, addressed in this Letter, is to identify the equivalence classes of states that are \emph{reversibly interconvertible} in the limit of asymptotically-many copies of the resource 
and to determine the rates of interconversion.  We show that all athermal states are reversibly interconvertible asymptotically and that the interconversion rate is governed by the free energies of the states involved.

The great merit of the resource theory approach is its generality. 
Rather than considering the behavior of the property of interest for some particular system with particular dynamics (as is typical in thermodynamics), one considers instead the fundamental limits that are imposed by the restriction defining the resource and \emph{the laws of quantum theory}. 
On the practical side, a better understanding of a given resource helps determine how best to implement the tasks that make use of it, and, more fundamentally, such an understanding may serve to clarify what sorts of resources are even relevant for a given task. For instance, entanglement is commonly asserted to  be the necessary resource for tasks in which the use of quantum systems yields improved performance over the use of classical systems. But in quantum metrology it is asymmetry which is relevant, not entanglement. 

Finally, the resource theory approach provides a framework for organizing and consolidating the
results in a given field, thermodynamics being particularly in need of such a framework, as well as synthesizing new results. 
Indeed, studying the interconvertibility of \emph{finite} resources leads to useful notions of free energy in that case, as shown by two of us in~\cite{horodecki-singleshot}, and to a more detailed, quantitative treatment of the second law, by three of us in~\cite{brandao_second_2013}.
Results similar to the former were also reported by {\AA}berg~\cite{aberg_truly_2011} and Egloff \etal~\cite{egloff_laws_2012}, who investigated the work extractable from finite resources.


\emph{Allowed Operations \& Resource States.---}We now define the restricted class of operations and the resource states more precisely. \ Given a
quantum system with Hilbert space $\mathcal{H}$ and Hamiltonian
$H$, the restricted operations are the completely-positive trace-preserving (CPTP) maps
$\mathcal{E}:\mathcal{L}\left(  \mathcal{H}\right)  \rightarrow
\mathcal{L}\left(  \mathcal{H}\right)  $ of the form%
\begin{align}
\mathcal{E}\left(  \rho\right)  =\mathrm{Tr}_{2}\left(  V_{12}\left(  \rho_1
\otimes\bar{\gamma}_2\right)  V_{12}^{\dag}\right),
\end{align}
where $\bar{\gamma}$ is the thermal (Gibbs) state of an arbitrary ancillary system with Hamiltonian $\bar{H}$ at inverse temperature $\beta=1/k_{\rm B}T$, and $V_{12}$ is an arbitrary unitary operation on the joint system which commutes with the total Hamiltonian: $[V_{12},H\otimes I+I\otimes \bar{H}]=0$. 
 Observe that $\mathcal E(\gamma)=\gamma$, where $\gamma$ is the Gibbs state associated with $H$. Any other state $\rho\neq \gamma$ is a resource state. While we here consider the case that input and output systems and their Hamiltonians are identical, this framework
can be easily extended to the more general case, as done by Janzing \etal~\cite{janzing_thermodynamic_2000}. 

The allowed operations are particularly relevant for thermodynamics because they cannot, on their own, be used to do work.
Moreover, it is not too difficult to see that various different kinds of athermal states \emph{can} be used, via the restricted class of operations, to do work: for thermal states at a temperature $T^{\prime}$ distinct from $T$ (hence athermal relative to $T$), work can be drawn using a heat engine (such states simulate having a second heat bath at a different temperature); for pure states within a degenerate energy eigenspace,  work can be drawn using a Szilard engine \cite{oppenheim_thermodynamical_2002}; for pure energy eigenstates, work can be drawn directly by an energy-conserving unitary.  One is led to expect that work can be extracted from \emph{any} athermal state.  We shall show that asymptotically this is indeed the case.

It is important to note the differences between the resource theory framework and the more usual approaches to thermodynamics. Chiefly, all sources and sinks of energy and entropy must be explicitly accounted for: only energy and entropy-neutral operations on the system and thermal reservoir are allowed, rather than specific energy- or entropy-changing operations more common in an open-system approach. All interactions between the system and reservoir are due to the unitary $V$ and not an interaction term in the total Hamiltonian. Moreover, no attempt is made \emph{a priori} to restrict the allowed operations to be physically realistic; indeed we assume the experimenter has complete control over $V$. This ensures that the restrictions we find are truly fundamental, though ultimately the operations needed to establish our main result are mappings between macroscopic observables and do not require fine-grained, microscopic control. These apparent differences notwithstanding, we show in the \sm that a number of different classes of operations for thermodynamics are in fact equivalent.

{\emph{Resource Interconvertibility \& Free Energy}.---}A central question in any resource theory is that of resource interconversion:  Which resources can be transformed into which others, and how easily? Generally there exists a partial order, or \emph{quasiorder}, of resources: We say $A\geq B$ if resource $A$ can be transformed into $B$ using the allowed operations. Functions which respect this quasiorder are known as \emph{resource monotones}. For instance, the relative entropy of entanglement is a well-known resource monotone relative to local operations and classical communication (LOCC)~\cite{horodecki_quantum_2009}.

Here we are interested in determining the optimal {rate} $R(A\,{\rightarrow}\,B)$ at which resource $A$ can generate resource $B$, in the limit of an infinite supply of $A$, that is, the largest $R$ such that $A^{\otimes n}\geq B^{\otimes nR}$ for $n\rightarrow \infty$.  A simple argument, going back to Carnot~\cite{fermi_thermodynamics_1956}, implies that if the transformation is reversible in the sense that $R(B\rightarrow A)\neq 0$, then the rate at which two resources can be \emph{reversibly} interconverted must achieve the optimal rate. Otherwise, it would be possible to generate arbitrary amounts of a resource state from a small number via cyclic transformations to and from another resource state.

That reversible interconversion is optimal (when possible) gives a simple means of characterizing the interconversion rate by using a ``standard'' reference resource. 
Consider a transformation from $A$ to $B$ which proceeds via the standard resource $C$: $A\rightarrow C\rightarrow B$. Following this with $B\rightarrow A$ must give a combined transformation of unit rate, again to avoid the possibility of spontaneously generating resources. Composing the rates, we have $R(A\rightarrow C) R(C\rightarrow B)R(B\rightarrow A)=1$, or
\begin{align}
R(A\rightarrow B)=\frac{R(A\rightarrow C)}{R(B\rightarrow C)},
\end{align}
using the fact that $R(A\rightarrow B)R(B\rightarrow A)=1$.
With this framework, we need only define the relative entropy $\rent{\rho}{\gamma}={\rm Tr}[\rho(\ln\rho-\ln\gamma)]$ to  state the main result of this Letter.
\begin{theorem}
\label{thm:rate}
Using thermal operations at background temperature $T$, asymptotic interconversion at nonvanishing rate is possible between all states $\rho$ and $\sigma$ of a system with Hamiltonian $H$. For $\gamma$ the Gibbs state of temperature $T$ associated with $H$, the optimal rate is given by
\begin{align}
\label{eq:conversionrate}
R(\rho\rightarrow \sigma)=\frac{\rent{\rho}{\gamma}}{\rent{\sigma}{\gamma}}.
\end{align}
\end{theorem}

Simple calculation reveals that $\rent{\rho}{\gamma}=\beta F_\beta(\rho)-\beta F_\beta(\gamma)$, where 
$F_\beta(\rho)\equiv \langle H\rangle_\rho-k_B T S(\rho)$ is the free energy and $S(\rho)=-\tr[\rho\ln \rho]$ the von Neumann entropy. Thus, the free energy directly determines the optimal rate of resource interconversion in our resource theory. 

To prove the result we shall employ the connection to free energy by constructing protocols for both \emph{distillation} of resource states into a standard state and \emph{formation} of resource states from standard states. The standard state is chosen to have energy but no entropy, so as to represent available work.

Before doing so, it is enlightening to note that, assuming reversible interconversion is possible, Eq.~(\ref{eq:conversionrate}) follows easily from \cite[Theorem 1]{horodecki_are_2002}, \cite[Theorem 4]{gour_measuring_2009}. This result states that any asymptotically-continuous resource monotone $f$ determines the interconversion rate via its regularization  $f^\infty(\rho)=\lim_{n\rightarrow\infty}\frac1nf(\rho^{\otimes n})$ as $R(\rho\rightarrow\sigma)=f^\infty(\rho)/f^\infty(\sigma)$, provided the latter is nonzero and finite. 

Here, $f(\rho)=\rent{\rho}{\gamma}$ is an athermality monotone (i.e.\ for all thermal operations $\mathcal{E}$, $\rent{\mathcal{E}(\rho)}{\gamma} \le \rent{\rho}{\gamma}$) by contractivity of the relative entropy under quantum operations and the fact that $\mathcal{E}(\gamma)=\gamma$. Its regularization is nonzero and finite since $f^\infty(\rho)=f(\rho)$, which follows from the additivity of the relative entropy and the fact the thermal state of $n$ identical systems is just $n$ copies of the thermal state of one system.  Finally, asymptotic continuity follows from  extensivity of energy by using Proposition 2 of~\cite{horodecki_locking_2005}; we leave the simple derivation of this to the \sm.  

Extensivity is crucial to the conclusion. For instance, $\tilde f(\rho)=\rent{\gamma}{\rho}$ (note the reversed order of $\rho$ and $\gamma$) is also an athermality monotone, but does not lead to the interconversion rate; the extensivity argument fails and $\tilde f$ is not asymptotically continuous. Nonetheless, $\tilde f(\rho)$ plays an important role in determining the resource requirements for creating low-temperature states~\cite{janzing_thermodynamic_2000}.

\emph{Distillation and Formation Protocols.---}In order to establish Theorem~\ref{thm:rate}, let us now turn to the distillation and formation protocols. For purposes of exposition, we specialize to the case of resources having just two nondegenerate energy levels, call them $\ket{0}$ and $\ket{1}$, i.e.\ qubits. This nevertheless captures the essential aspects of the problem. We first consider the distillation and formation of \emph{quasiclassical} resources $\rho$, meaning $[\rho,H]=0$ and take up the case of non-stationary resources afterwards. 
In what follows we sketch the steps required to complete the proof and leave the somewhat cumbersome mathematical details to the \sm. 
 
Both distillation and formation protocols must satisfy three requirements, up to error terms smaller than $O(n)$: (1) energy conservation, (2) unitarity, and (3) equality of input and output dimensions. Without loss of generality, we may take the total Hamiltonian to be $H=E_0\sum_i \ket{1}_i\bra{1}$ for some energy $E_0$, where the sum runs over all the qubits.

We begin the distillation protocol with $\ell$ copies of the Gibbs state $\gamma$ of $H$ and $n$ copies of the resource $\rho$, where $\rho=(1-p)\ketbra{0}+p\ketbra 1$ for arbitrary $0\leq p\leq 1$ and $\gamma=(1-q)\ketbra 0+q\ketbra 1$ for $q=e^{-\beta E_0}/(1+e^{-\beta E_0})$. The aim is to effect a transformation of the form $
\gamma^{\ot \ell} \ot \rho^{\ot n} \to \sigma^{(k)}\ot \ket{1}\bra{1}^{\ot m}
$ 
by an energy-conserving unitary, such that $m$ is as large as possible. The resulting \emph{exhaust} state $\sigma$ of $k$ systems is arbitrary, though as an aside we show that the optimality of the protocol implies that it has near-Gibbs form in the \sm. 
We denote by $R=\frac{m}{n}$ the rate of distillation and
$\ep=\frac{n}{l}$ the ratio between initial resource states and Gibbs states.
The Gibbs states are free, so we allow $\ep\rightarrow 0$  as $n\rightarrow\infty$.

We now use the fact that for large $n$, $\rho^{\ot n}$ consists of mixtures of basis states corresponding to length-$n$ binary strings with roughly $np$ 1s. The number $t$ of 1s in a string is known as its \emph{type}, and more concretely we have that, to an error which vanishes as $n\rightarrow\infty$,
\begin{align}
\rho^{\ot n} \approx \sum_t p_t P_t.
\label{eq:typesupport}
\end{align}
Here the $t$ summation runs over strongly typical types, the types for which $t=np\pm O(\sqrt{n})$~\cite{cover_elements_2006},
and $P_t$ denotes the projector onto the type $t$. An entirely similar statement holds for $\gamma^{\ot \ell}$.
For simplicity we shall first pretend that both $\gamma^{\ot \ell} $
and $\rho^{\ot n}$ consist of a single type and subsequently show how to extend the argument to all strongly-typical types.

We begin with a single composite type, a concatenation of a type coming from the resource state and one from the reservoir state. This corresponds to a uniform mixture of strings of length $n+\ell$, each of which consists of two substrings: the first having $\ell q$ 1s and the second $np$ 1s. There are roughly $e^{l \h2(q)}\times e^{n \h2(p)}$ such strings, where $\h2(p)=-p\ln p-(1-p)\ln(1-p)$ is the binary entropy, expressed in nats.

Now consider a transformation which maps these strings to new strings having at least $m$ 1s in the rightmost positions,
\begin{align*}
\overbrace{00 \dotso 0\underbrace{11 \dotso 1}_{\ell q}}^\ell\,
\overbrace{00 \dotso 0\underbrace{11 \dotso 1}_{np}}^n \,\to\, \underbrace{0 \dotso 00\,\underbrace{1 \ldots 1}_{rk}}_{k}\,
\underbrace{1 \ldots 1}_{m}
\end{align*}
where $k=\ell+n-m$ expresses {\it conservation of dimension}, and $r$ and $m$ are to be determined.
{\it Conservation of energy} requires that the number of 1s  is conserved, hence
$\ell q+ np=rk+m$.
{\it Unitarity} requires that
there are at least as many strings of length $k$
with $rk$ 1s as the number of initial strings: $e^{k \h2(r)}\geq e^{\ell \h2(q)+n \h2(p)} $.
Roughly speaking, this is {\it conservation of entropy}.
Using these three conditions, we find that the transformation is possible for any $R$ such that
\begin{align}
\h2(q) + \ep \h2(p)\leq (1+ \ep -R \ep)\,\h2\left(\frac{q+ \ep p -\ep R}{1 +\ep -\ep R}\right).
\end{align}
We now expand this with respect to $\ep$ to
first order and let $\ep\to 0$.
This means 
the heat reservoir is much larger than the resource systems.
As a result we obtain that the following rate can be achieved
\begin{align}
R= \frac{\h2(q)-\h2(p)+\beta (p-q)}{\h2(q)+\beta(1-q)}=\frac{\rent{\rho}{\gamma}}
{\rent{\ketbra{1}}{\gamma}},
\label{eq:R}
\end{align}
establishing one direction necessary for Theorem~\ref{thm:rate}.

In the above argument we worked with a single composite type, whereas in actuality the initial state is a mixture of these. 
We thus apply the protocol separately to each composite type, 
assuming the number $m$  of output excited states to be the same
for all input types, with $m$ fitted to
the composite type containing the fewest strings (i.e.\ the one consisting of strings with $\ell q-O(\sqrt{\ell})+ np-O(\sqrt{n})$ 1's).
To proceed as above we need to ensure that any variations from the above conditions are small relative to $n$. Thus, we need to simultaneously fulfill $\sqrt{\ell}\ll m=Rn$, in order for $R$ from \eqref{eq:R} to be achievable, and
$\ell \gg n$, in order for $\epsilon\rightarrow 0$. Choosing $\ell=(Rn)^{3/2}$ therefore ensures that our estimate \eqref{eq:R} will be accurate in the limit $n\rightarrow \infty$.~\footnote{Additionally, the number of strings of a given type $np$ is given by poly$(n)e^{n\h2(p)}$, not simply $e^{n\h2x(p)}$. However, in the limit $n\rightarrow \infty$, the polynomial factor is again not relevant in our estimates.}

The formation protocol is similar to the distillation protocol and is again based on considering type transformations satisfying the three requirements of energy conservation, unitarity, and dimension conservation. The major difference is that whereas the ideal distillation output is simply the fixed-type state $\ket{1}^{\ot m}$, the ideal formation output must recreate a good approximation to the probabilistic mixture of type classes found in $\rho^{\ot n}$.

We construct the formation protocol in three stages. The first two are similar to the distillation protocol. In the first, a given type class of the Gibbs state together with the standard resource is transformed into a desired type class of the target resource $\rho^{\ot n}$. In the second, the transformation is extended to all the strongly-typical types of the Gibbs state. Finally, in the third step an additional number of Gibbs states are used to probabilistically select which type class of the target should be output, in order to recreate the appropriate distribution over types of the target state. In principle this step is irreversible, but since the number of type classes grows only polynomially with $n$, the number of extra resources required for the third step of the formation protocol vanishes in the $n\rightarrow \infty$ limit. The similarity of the first two steps with the distillation protocol then ensures that the formation protocol achieves the inverse rate.

Distillation for arbitrary resource states is related to that of stationary states, and we can recycle part of the previous distillation protocol. Suppose the resource state has the diagonal form $\rho=p\ketbra{\phi_1}+(1-p)\ketbra{\phi_2}$, for arbitrary orthogonal states $\ket{\phi_k}$, implying an average energy of $\langle E\rangle=(p|\bracket{\phi_1}{1}|^2+(1-p)|\bracket{\phi_2}{1}|^2)E_0$. In $n$ instances of $\rho$ the total energy will overwhelmingly likely be $n\langle E\rangle\pm O(\sqrt{n})$. Now imagine projecting the resource state onto the various energy subspaces, destroying any coherence between them.
Just as in \eqref{eq:typesupport}, $\rho^{\ot n}$ is supported almost entirely on its typical subspace, whose size not larger than $e^{nS(\rho)+O(\sqrt{n})}$. Thus, the state support in every energy subspace is at most this large.

Now we may imagine applying the same scheme as in the previous distillation protocol, creating as many copies of $\ket{1}$ as possible. The three conditions now become $k=\ell+n-m$, $\ell qE_0+n\langle E\rangle=rkE_0+mE_0$, and $e^{k\h2(r)}\geq e^{\ell \h2(q)+nS(\rho)}$. An entirely similar derivation leads again to the distillation rate found in \eqref{eq:R}. Finally, since the distillation operations commute with the Hamiltonian, they commute with the projection onto energy subspaces. Thus we may instead imagine that this projection is performed \emph{after} the distillation step. Such a projection has no effect on the work systems, while the form of the exhaust state is irrelevant, and therefore we may dispense with the projection step altogether.

The formation of arbitrary resource states is more complicated than their distillation. Strictly speaking, the desired transformation is impossible, since the inputs are states diagonal in the energy basis and the allowed transformations cannot change this fact. However, to create the appropriate coherences between energy subspaces it suffices to use a small additional resource in the form of a superposition over energy eigenstates. 

In particular, a system in a superposition of energy levels acts as  a reference system which lifts the superselection rule of energy conservation, as in \cite{aharonov-susskind,bartlett_reference_2007}, allowing one to create arbitrary coherences over energy levels on the system. However, since $\rho^{\ot n}$ is almost entirely supported on energy levels in the range  
$n\langle E\rangle\pm O(\sqrt{n})$,  the formation process requires only a reference system made from order $\sqrt{n}$ qubits. The extra resource of the reference system is thus of a size sublinear in $n$ and does not affect the rate calculations.  
This creates an interesting asymmetry between distillation and formation, akin to a similar phenomenon in the resource theory of entanglementmt, where distillation of entangled states does not require any communication but formation requires an amount sublinear in the number of inputs $n$.

\emph{Conclusions}.---We have shown that well-known results from thermodynamics can be derived quite naturally within the framework of the resource theory of energy-preserving transformations and auxiliary thermal states. We should emphasize that although the procedures we have described for the conversion of resource states may seem quite unnatural from a physical point of view, their use is to establish the ``in principle'' interconversion rate given in Theorem~\ref{thm:rate}. Any more realistic reversible transformation, for instance the Hamiltonian method of~\cite{alicki_thermodynamics_2004} or the sequential protocol (for quasiclassical resources) of~\cite{skrzypczyk_extracting_2013}, will necessarily extract the same amount of work.

\emph{Acknowledgments}.---We thank Jochen Rau and Dominik Janzing for helpful conversations. JMR acknowledges support from the Center for Advanced Security Research Darmstadt (CASED). RWS acknowledges support from the Government of Canada through NSERC and the Province on Ontario through MRI. MH thanks the support by Foundation for Polish Science TEAM project 
cofinanced by the EU European Regional Development Fund for preparing the final version of this paper.
Part of this work was done at National Quantum Information Centre of Gdansk. The authors thank the hospitality of Institute Mittag Leffler within the program Quantum Information Science (2010), where part of this work was done.

\bibliographystyle{apsrev4-1}
\bibliography{work}

\onecolumngrid
\clearpage

\widetext
\appendix
\setcounter {equation} {0} 
\begin{center}
	{\bf \small {APPENDIX}}
\end{center}
This appendix contains eight sections. The first shows that the relative entropy distance to the Gibbs state is an asymptotically continuous function. The next four sections discuss in detail the state transformation protocols for the case of two-level systems. Section II presents a distillation protocol for quasi classical states, while section III describes a formation protocol also for quasi classical states. The following two sections extend these protocols to the case of arbitrary nonstationary two-level resources. Then in Section VI we outline how the results can be easily generalized to higher dimensions by considering as an example the distillation protocol for quasi classical states.  Section VII discusses some characteristics of the exhaust states produced in these protocols. Finally, Section VIII discusses the equivalence of our formulation to other models of thermodynamics and the degree of control one needs to implement our thermal operations.

\section{Extensivity and asymptotic continuity}
To show that $\rent{\rho}{\gamma}$ is asymptotically-continuous, we make use of the following, from~\cite{horodecki_locking_2005+}:
\begin{proposition}
Suppose a function $f$ satisfies (1) ``approximate affinity''
\begin{align}
|pf(\rho)+(1-p)f(\sigma)-f(p\rho+(1-p)\sigma))|\leq c,
\end{align}
for some constant $c>0$ and any $p$ such that $0\leq p\leq 1$, and (2) ``subextensivity'' $f(\rho)\leq M\log d$, where $M>0$ is constant and $d={\rm dim}(\mathcal{H})$ for $\mathcal{H}$ the state space on which $\rho$ has support. Then $f$ is asymptotically continuous:
\begin{align}
|f(\rho_1)-f(\rho_2)|\leq M\| \rho_1-\rho_2\|_1\log d+4c.
\end{align}
\end{proposition}

The entropy relative to the Gibbs state, $f(\rho):=\rent{\rho}{\sigma}$, satisfies both conditions. To see the first, let $\tau=\sum_k p_k\tau_k$ for some arbitrary set of density operators $\{ \tau_k \}$ and probability distribution $p_k$ and let $\omega$ be another arbitrary density operator. Then
\begin{align} 
\rent{\tau}{\omega}&={\rm Tr}\left[\sum_k p_k\left(-\tau_k\log\omega+\tau_k\log\tau\right)\right]\\
&={\rm Tr}\left[\sum_k p_k\left(\tau_k\log\tau_k-\tau_k\log\omega+\tau_k\log\tau-\tau_k\log\tau_k\right)\right]\\
&=\sum_k p_k \rent{\tau_k}{\omega}+\sum_k p_k S(\tau_k)-S(\tau),
\end{align}
where in the second line we have added and subtracted ${\rm Tr}(\tau \log \tau_k)$. Since $S(\tau)\leq \sum_k p_kS(\tau_k)+H(p_k)$ where $H(p_k)$ denotes the Shannon entropy for the distribution $p_k$, and since the relative entropy is convex, this implies 
\begin{align}
0\leq \sum_k p_k \rent{\tau_k}{\omega}-\rent{\tau}{\omega}\leq H(p_k).
\end{align}
Finally, letting $\omega=\gamma$ and $\{ \tau_k\} = \{\rho,\sigma\}$ with distribution $(p,1-p)$, we find
\begin{align}
p\rent{\rho}{\gamma}+(1-p)\rent{\sigma}{\gamma}-\rent{p\rho+(1-p)\sigma}{\gamma}\leq h_2(p)\leq 1,
\end{align}
where $h_2$ is the binary entropy function. 

The fact that the second condition is satisfied, i.e., that the entropy relative to the Gibbs state is subextensive, follows from the fact that
the maximum energy of the system is extensive. First, note that $\rent{\rho}{\gamma}=\beta F_\beta(\rho)-\beta F_\beta(\gamma)$, where $F_\beta(\rho)=\langle H\rangle_\rho-\frac{1}{\beta}S(\rho)$ is the free energy. Thus, the maximum of $\rent{\rho}{\gamma}$  occurs for $\rho=\ketbra{E_{\rm max}}$ where $\ket{E_{\rm max}}$ is the eigenstate of maximum energy. Direct calculation shows
\begin{align}
\rent{\ketbra{E_{\rm max}}}{\gamma}&={\rm Tr}\left[\ketbra{E_{\rm max}}\left(\log \ketbra{E_{\rm max}}-\log \gamma\right)\right]=-\bra{E_{\rm max}}\log\gamma\ket{E_{\rm max}}\\
&=\beta E_{\rm max}+\log Z_\beta=\beta E_{\rm max}+\log {\sum_k e^{-\beta E_k}}\leq \beta E_{\rm max}+\log d.
\end{align}
Here we have assumed that the energy values $E_j>0$. When the maximum energy is extensive, i.e.\ $E_{\rm max}\leq K\log d$ for some constant $K$, we obtain $\rent{\ketbra{E_{\rm max}}}{\gamma}\leq M\log d$ for $M=\beta K+1$.

\section{Distillation of quasiclassical states}

For simplicity of presentation, we consider qubit systems with the Hamiltonian given by $H=\sum_i \ket{1}_i \bra{1}$, where the sum runs over all involved qubits.   We start with $l$ copies of the Gibbs state $\gamma$ and $n$ copies of the resource state $\rho$, where
\be
\rho=(1-p) \ket{0}\bra{0}+p \ket{1}\bra{1};\quad 
\gamma=(1-q) \ket{0}\bra{0}+q\ket{1}\bra{1},
\ee
with $q=e^{-\beta}/(1+ e^{-\beta})$ and $\beta$ the inverse temperature, which we take as a constant parameter. The aim is to obtain the maximal number of copies possible of qubits in the pure excited state $\ket{1}$ by implementing a unitary that commutes with $H$ and taking the partial trace over some subsystem,
\be
\gamma^{\ot l} \ot \rho^{\ot n} \to \sigma^{(k)}\ot \ket{1}\bra{1}^{\ot m}.
\ee

We denote $R=\frac{m}{n}$ (the rate of distillation) and 
$\ep=\frac{n}{l}$ (the ratio between the number of used Gibbs states 
and the number of resource states). The Gibbs states are free, so 
we accept that $\ep$ asymptotically vanishes. 

In the protocol we shall use the fact that up to a small error (vanishing for a large number of qubits) 
\be
\rho^{\ot n} \approx \sum_t p_t P_t
\ee
where $t$ run over strongly typical types,
i.e.\ the types containing strings with the number of $1$'s 
within the interval  $(n p - O(\sqrt{n}), np +O(\sqrt{n}))$,
and $P_t$ denotes the projector onto type $t$. Similarly 
\be
\gamma^{\ot l}\approx \sum_t q_t Q_t
\ee
again with $q_t\approx qn$. The errors in both approximations are smaller than $2^{-\sqrt{n}}$ when quantified by the trace norm.

For simplicity we shall first pretend that both $\gamma^{\ot l} $ and $\rho^{\ot n}$ consist of a single type. Then further we will show how to extend the argument to a mixture of types.

So we start with a tensor product of two types (one from Gibbs, 
the other from $\rho$), i.e.\ an equal mixture of strings of length $l+n$. The string consist of two substrings:
the first has $ql$ $1$'s and the second has $pn$ $1$'s:
\be
\overbrace{000 \ldots 0\,\underbrace{11 \ldots 1}_{lq}}^l\, 
\overbrace{00 \ldots 0\,\underbrace{111 \ldots 1}_{np}}^n.
\ee
There are roughly 
\be
2^{l \h2(q)}\times 2^{n \h2(p)}
\label{eq:entpq}
\ee 
such strings (with the error being a multiplicative $\text{poly}(n)$ factor).

We now apply a unitary transformation to these strings to map them into strings of the same total length which have $m$ $1$'s
to the right:
\begin{eqnarray}
&& \overbrace{000 \ldots 0\,\underbrace{11 \ldots 1}_{lq}}^l\, 
\overbrace{00 \ldots 0\,\underbrace{111 \ldots 1}_{np}}^n \quad\to\quad
\underbrace{00000 \ldots 0000\,\underbrace{111 \ldots 11}_{rk}}_{k}\, 
\underbrace{111 \ldots 1}_{m}
\end{eqnarray}
where 
\be
k=l+n-m
\label{eq:con-dim}
\ee (conservation of dimension), and $r$ and $m$ are about to be determined. 
First, $r$ 
is fixed by {\it conservation of energy}, which requires that the number of $1$'s  is conserved:
\be
lq+ np=rk+m.
\label{eq:con-en}.
\ee
Then {\it unitarity} requires that 
there are at least as many strings of length $k$ 
with $rk$ $1$'s as the number of initial strings \eqref{eq:entpq}:
\be
2^{k \h2(r)}\geq 2^{l \h2(q)+n \h2(p)}.
\label{eq:con-ent}
\ee
Roughly speaking this is {\it conservation of entropy}. 
Using \eqref{eq:con-en} and \eqref{eq:con-ent} we obtain 
that our transformation is possible if 
\be
\h2(q) + \ep \h2(p)\leq (1+ \ep -R \ep)\,\h2\left(\frac{q+ \ep p -\ep R}{1 +\ep -\ep R}\right),
\ee
where recall that $\ep$ is the ratio of Gibbs states used,
and $R$ the ratio of pure excited states obtained. 
We now expand this with respect to $\ep$ to  first order and let $\ep\to 0$. This means that we take many more Gibbs states than the resource states.  In other words, we presume a heat reservoir that is arbitrarily larger than the size of our system. As a result, recalling that $q=e^{-\beta}/(1+ e^{-\beta})$, we obtain that the following rate can be achieved
\be
R=\frac{\h2(q)-\h2(p)+\beta (p-q)}{\h2(q)+\beta(1-q)}=\frac{S(\rho || \gamma)}
{S( \ket{1}\bra{1} || \gamma)},
\ee
i.e., the protocol achieves the upper bound obtained by the monotonicity argument in the main text.

So far we have worked with a single type. But our initial state is actually a mixture of products of types $Q_t \ot P_{t'}$.
We thus apply the protocol separately to each type,
with the same number $m$ of required output excited states,
for all types, with this number being fitted to 
the product of less numerous types (i.e.\ the one with the smallest number of $1$'s, namely $(np-O(\sqrt{n}))(lq-O(\sqrt{l}))$ $1$'s). 
Note, however, that the variations will disappear asymptotically, as we divide the equations by $l$.
Also the approximation to the number of strings in each type by the exponential of the entropy is correct up to a multiplicative polynomial factor, which is also irrelevant asymptotically.

\section{Formation of quasiclassical states}

We are going to construct the formation protocol in three stages.  
The first is to show that for a particular type $T_q$ of the $l$ copies of the Gibbs state, we can
create a particular type $T_p$ of the state we want to form.  We then show
that we can do this for all types of the Gibbs state.  Finally we show
how to correctly get the distribution over types of the target state.

We shall need the following useful lemma:
\begin{lemma} 
\label{lemma:birkoff}(Birkhoff primitive) The following operation can
be done by means of thermal operations with arbitrary accuracy: 
\be
\rho \to \sum_k p_k U_k \rho U_k^\dagger
\label{eq:bir}
\ee
where $p_k$ is an arbitrary probability distribution, 
and the $U_k$ commute with the Hamiltonian. In particular, 
a random permutation of the systems is a valid operation. 
\end{lemma}
{\bf Remark.} The accuracy depends on 
the number of Gibbs states that are used, but in our paradigm they are for free. 
{\bf Proof.} First, let us note that the following unitary transformation preserves energy:
\be
\sum_i \ketbra{i}\ot \tilde U_i
\ee
provided that $\ket{i}$ are eigenvectors of the reservoir Hamiltonian and the $\tilde U_i$ commute with the system Hamiltonian. To obtain the required transformation, we take the initial state of the reservoir to be $l$ copies of the Gibbs state. Let $q_i$ denote the probability distribution of single strings.
We now divide the set of eigenvectors of the reservoir Hamiltonian into sets, denoted $S_k$, such that the sum of the probability distribution over $i$   within each set yields approximately the probability $p_k$ from \eqref{eq:bir}, that is, 
$\sum_{i\in S_k} q_i \approx p_k$.
This can be done with arbitrary accuracy by taking $l$ large enough, since $q_i\leq \max\{q,1-q\}^l$.
Then, for every $i\in S_k$, we set  $\tilde U_i=U_k$. 
\hfill$\square$

We want to form $n$ copies of the state 
$\rho=(1-p) \ketbra{0}+p\ketbra{1}$
from a pure excited state. Let us first show how we can form any of 
the typical types of the state.  

\vspace{0.2 cm}

{\bf Formation of a maximally mixed state over a fixed type.} 

\vspace{0.2 cm}

Consider a fixed type $T_p$  of the state of $n$ systems that we want to create, and let 
it be one with $np$ $1$'s. For other types, the reasoning is the same and the asymptotic 
rate will be the same. We start with $l$ copies of the Gibbs state $\gamma$, and $m$ 
copies of the excited state $\ket{1}$.  We consider a final exhaust system
consisting of $k$ two-level systems.  
Consider a typical type $T_q$ of the Gibbs state; it has $lq$ $1$'s.  
In that type, there are $\approx 2^{l \h2(q)}$ strings up to some  $2^{\sqrt{l} \h2(q)}$ factor.  We want to map these initial strings onto the $N$ final strings in the type $T_p$
as follows.  Take $\{u_i\}_i$ to be the set of strings in  $T_p$ and for each
string $u_i$ consider some set $\{v^i_j\}_j$ of strings on the exhaust system.  We
now map each of the initial strings to some string $u_iv^i_j$. This is illustrated in Figure~1.

\begin{figure}
		\includegraphics[width=8cm]{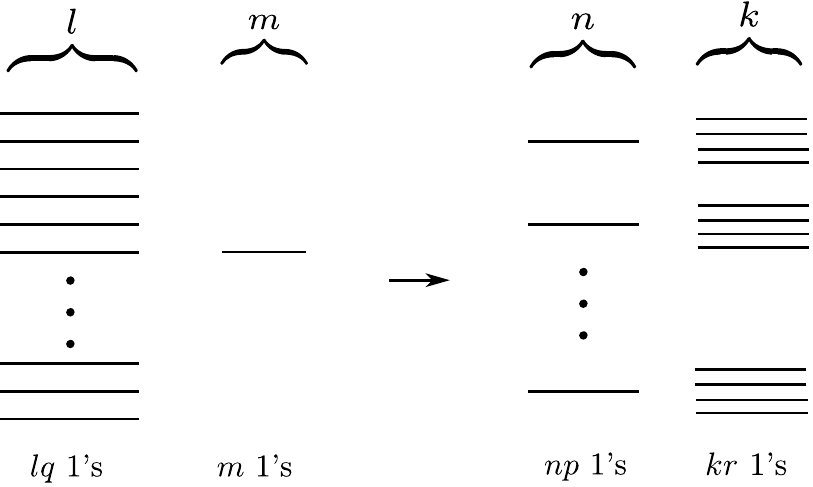}
	\label{fig:formation}
		\caption{Mapping of strings in the formation protocol}
\end{figure}

For sets $\{v^i_j\}_j$ corresponding to different values of $i$, we can take the number of  strings in each set  to be the same
or off by $1$, simply by assigning the strings in an order determined
by fixing $j=1$, then incrementing the $i$ register until $i=N$, then incrementing
the $j$ register by $1$, reseting $i$ to $1$ and again incrementing the $i$ register until $i=N$ and repeating.  This is all done to ensure that when we
trace out the exhaust system, we get an even mixture over permutations
within the type class.

We can now use the analogues of equations
(\ref{eq:con-dim}), (\ref{eq:con-en}) and (\ref{eq:con-ent})
to ensure that we can perform the unitary which implements this mapping:
\be
m+l=n+k
\label{eq:revcon-dim}
\ee
\be
lq+ m=rk +np
\label{eq:revcon-en}
\ee
\be
2^{l \h2(q)}  \leq 2^{k \h2(r)+n \h2(p)}
\label{eq:revcon-ent}
\ee

We now take $l\propto m^\alpha$ with $1<\alpha<2$ and do as before.  We take $\alpha>1$ so that we can 
take  $\epsilon=n/l$ small, and $\alpha<2$ so that $\sqrt{l}$ is sublinear
in $m$ and we can ignore such terms.  

This maps type $T_q$ of the $l$ Gibbs states onto the type $T_p$ of $\rho^{\otimes n}$.
We can map each of the initial Gibbs types onto $T_p$
in this manner using a unitary $U_{pq}$.  For each such mapping, the above three equations will change,
but only by some $\sqrt{l}$ factors which we took to be sublinear in $m$.  
We can thus choose $m$ to ensure conservation of energy in the worst case of Eq. (\ref{eq:revcon-en}),
and $k \h2(r)$ is chosen to ensure the inequality Eq. (\ref{eq:revcon-ent}) in
the worst case.

We now need to implement $U_{pq}$ conditioned on the initial type $T_q$.
Since the typical initial types are on orthogonal and diagonal subspaces, 
we first do a conditional copying of the type class $q$ onto an initialised
register. We then act $U_{pq}$ conditioned on this register. 
Since the number of types is polynomial in $l$, the
register only needs to be of size $\log{l}$, and thus this 
resource does not matter as it is sublinear in $m$. It is an interesting question whether the formation 
protocol can be made to work without this sublinear supply of pure states.

We denote by $U_p$ the unitary that creates a particular type $T_p$. To get the distribution
over types, we simply use the Birkoff primitive of Lemma 
\ref{lemma:birkoff} to implement $\sum_p \ket{p}\bra{p} \otimes U_{p}$.  This is irreversible, but since the number
of typical types is polynomial in $n$, the rate of entropy that is created
by this procedure is negligible, i.e.\ logarithmic in $n$. It is not hard to see that Eqs. (\ref{eq:revcon-dim}-\ref{eq:revcon-ent}) give the required rate (i.e.\ the inverse of the distillation rate).

\section{Distillation of arbitrary states}

We now extend distillation to the case where our states are not
diagonal in the energy eigenbasis. 

Consider a state $\rho=p \ketbra{\phi_1}+(1-p)\ketbra{\phi_2}$.
The average energy of the state is $\mean{E}= p|\bracket{\phi_1}{1}|^2+ (1-p) |\bracket{\phi_2}{1}|^2$.
As before, we consider $n$ copies of $\rho$ and $l$ copies of systems in a Gibbs state $\gamma$. 
Regarding $\rho^{\ot n}$, only the blocks with energy $E\in [n \mean{E} - \sqrt{n}, n\mean{E} + \sqrt{n}]$ 
will be relevant, i.e.\ we have
\be
\tr \left( \sum_E P_E \rho^{\ot n} \right) \geq 1- 2^{-O(n)}
\ee
where the sum runs over $E \in [n \mean{E} - O(\sqrt{n}), n\mean{E} + O(\sqrt{n})]$,
with $P_E$ the projector onto the energy $E$ eigenspace
(this follows from Eq. \eqref{eq:tails-av-E}, proven in Section \ref{sec:form_arb}).

Our protocol has two stages: 
\begin{itemize}
\item[(i)] unitary rotation within energy blocks of a resource system (consisting of $n$ qubits) solely. 
\item[(ii)] drawing work by string permutations on the total system resource (with $n$ qubits) plus heat bath
(with $l$ qubits)
\end{itemize}
We write down the resource state in the energy eigenbasis. As said above, only blocks 
with energy $\approx nE$ will appear.  We will use the fact that 
the state is, up to exponentially vanishing error, equal to its projection onto the typical subspace, having 
dimension $2^{nS(\rho)+O(\sqrt{n})}$ where $S(\rho)$ is the von Neumann entropy of $\rho$. Therefore,  within every block the rank of the state is not larger than $\approx 2^{n S}$ (as a projection cannot increase the rank). 

Now stage $(i)$ is the following: within each energy block, we apply unitary rotation, which diagonalizes the state restricted to the block 
in the energy basis. Then there is stage $(ii)$, in which   
we apply the protocol of distillation of quasiclassical states as in Section \ref{sec:distq},
i.e.\ we permute strings in such a way that the output strings have $m$ $1$'s to the right. 
Such a protocol produces $m$ systems in a pure excited state (note that all coherences 
initially present in the state are now left in the garbage).
Using the same notation as in Section \ref{sec:distq},  for a product of two single types, e.g. 
with $lq$ and $n\mean{E}$ $1$'s, respectively, the constraints now become 
\be
&&k=m+m-l \nonumber \\
&&lq + n\mean{E}= rk +m\nonumber\\
&& 2^{k \h2(r)}\geq 2^{l \h2(q)+n S(\rho)} 
\ee
As before, taking the limit $\frac{n}{l}\to 0$ we obtain 
that any rate $R$ is achievable, provided it satisfies
\be
R\leq \frac{\h2(q)-S(\rho)+\beta (\mean{E}-q)}{\h2(q)+\beta(1-q)}=\frac{S(\rho ||\gamma)}
{S(\ketbra{1} || \gamma)}. 
\ee
Thus also for states that are not quasiclassical, we can reach the upper bound given by the relative entropy distance from the Gibbs state.

\section{Formation of arbitrary states}
\label{sec:form_arb}
We now show that we can achieve reversibility even in the case of states that are not quasiclassical.  To do so, however, we must allow the use of a sublinear amount of states that are a superposition over energy eigenstates.  This is a reasonable assumption since the rate at which such states are consumed vanishes in the asymptotic limit.  This is very similar
to the fact that in entanglement theory, distillation requires no communication
but formation requires a sublinear amount of it. Or, instead, formation requires a state which is a superposition over different amounts
of entanglement.  The superposition over different amount of entanglement (known as entanglement spread \cite{HP97, HW03, HL04, Har09}), 
is analogous to the superposition over energy eigenstates in the present context. In the athermality context, distillation of work requires no superposition over energy eigenstates, but formation does. 

Suppose we want to implement some unitary
\be
U=\sum_{ij}u_{ij}\ket{E_i}\bra{E_j}
\ee
that does not conserve energy. We introduce a state that acts as a reference frame for time,
\be
\ket{H}=\sum f(h)\ket{h},
\label{eq:qrefsys}
\ee 
where $\ket{h}$ denotes an energy eigenstate, and we implement
\be
U^{\textrm{inv}}=\sum_{ij}u_{ij}\ket{E_i}\bra{E_j}\otimes \ket{h-E_i+E_j}\bra{h}
\,\, 
\label{eq:energyswap}
\ee
on the system and reference frame.  If we are interested in implementing $U$ on the state
\be
\ket{\psi}=\sum_{i\in \cal{S}} c_i\ket{E_i},
\ee
then we do so by implementing $U^{\textrm{inv}}$ on $\ket{\psi}\otimes \ket{H}$.   Note that we must ensure that the reference frame system has energy levels with gaps of size $|E_i-E_j|$ for every transition appearing in $U$.  
If the energy spread of $f(h)$ is large compared to the largest value of $|E_i-E_j|$ in an energy transition induced by $U$, then the state of the reference frame is not disturbed
very much in the process.

For the problem in which we are interested, this is indeed the case because on the typical subspace, the variation in energy is sublinear in the number of copies of the state we want to create. To see how this works by way of example, note that if in Eq.~\eqref{eq:qrefsys}, we take $f(h)$ to be $1/\sqrt{N}$ for energies $h \in \{1,...,N\}$ and $f(h)=0$ otherwise,
then removing a unit of energy and adding it to another system, does not change the  state of Eq.~\eqref{eq:qrefsys} much. i.e.\ 
the inner product between $\ket{H}$ and $\sum_h \ket{h-1}\bra{h} \ket{H}$ approaches $1$, because
\be
\sum_{h=1}^N \frac{1}{\sqrt{N}} \bra{h}\sum_{h=0}^{N-1}\frac{1}{\sqrt{N}} \ket{h}=1-\frac{1}{N},
\ee
which approaches $1$ for large $N$.  States like that of Eq.~\eqref{eq:qrefsys} therefore allow us to lift the superselection rule for
energy, without being consumed much in the process.

This gives some insight into embezzling states~\cite{ent-embezzling}.  These are resource states that are often used in entanglement theory in similar situations.  For instance, one can use a state similar in form to Eq.~\eqref{eq:qrefsys}
\be
\ket{E}=\sum f(k)\ket{\phi}_{AB}^{\otimes k}\ket{00}_{AB}^{\otimes (n-k)}
\label{eq:embezzling}
\ee
which is a superposition of a different number of entangled EPR pairs $\ket{\phi}_{AB}$.  These 
states can be used to implement operations which
need to create superpositions over amounts of entanglement (entanglement spread).  Just as removing one unit of energy, doesn't change the state of Eq.~\eqref{eq:qrefsys} much, likewise,
removing one EPR pair from the state of Eq.~\eqref{eq:embezzling} and adding it to another system doesn't change the embezzling state by much.  We can embezzle energy, just as one can embezzle entanglement.  We therefore see that a superposition over some resource can create an embezzling state for that resource, and will allow us to lift some superselection rule or restriction.

With this small superposition over energy states, let us now show that we can create an arbitrary state at a rate given by the relative
entropy distance to the Gibbs state. Let $\rho := p \ket{\phi_1}\bra{\phi_1} + (1 - p) \ket{\phi_2}\bra{\phi_2}$ and
\begin{equation}
\rho^{\otimes n} = \sum_{k, g} p_k \ket{\Psi_{k, g}}\bra{\Psi_{k, g}},
\end{equation}
with
\begin{equation} \label{formPsi}
\ket{\Psi_{k, g}} := \pi_{g} \ket{\phi_1}^{\otimes k} \otimes \ket{\phi_2}^{\otimes n - k}, 
\end{equation}
for $\pi_g$ a permutation. 

The idea of the protocol is as follows: we will first create a diagonal
state  
\be
\varrho_n=\sum  p_k  \ket{{t_k, s_g}}\bra{{t_k, s_g}}
\ee which has the same spectrum as $\rho^{\otimes n}$ and
where each eigenstate has the same average energy as an eigenstate in the
typical subspace of  $\rho^{\otimes n}$. From the result
of the previous section it is not hard to see that 
this can be done at a rate given
by the relative entropy distance of $\rho$ to the Gibbs state, since in
the limit of many copies, the regularised relative entropy distance is the same.  We would then like to rotate
the diagonal basis to the $\Psi_{k, s}$-basis.  This cannot be done by 
unitaries which commute with the Hamiltonian unless we allow for a reference
frame $\ket{H}$ which is a superposition over energy states.  We then want to show
that the reference frame which allows us to break the energy superselection 
rule is consumed at a vanishingly small rate.  We do so by showing that
the reference frame superposition is over a size sublinear in $n$.  This
can be understood as coming from the fact that in the typical subspace,
the superposition over different types is sublinear.

We consider only typical $\ket{\Psi_{k, g}}$ with $k \in \text{Typ}_{\rho} := [np - \sqrt{n}, np + \sqrt{n}]$. Then
\begin{equation} \label{typpicalPsits}
\left \Vert  \rho^{\otimes n}  -    \sum_{k \in \text{Typ}_{\rho}, g} p_k \ket{\Psi_{k, g}}\bra{\Psi_{k, g}}  \right \Vert_1 \leq 2^{- \Omega(\sqrt{n})}
\end{equation}

Let also
\begin{equation} 
\label{lineardecompPsi}
\ket{\Psi_{k, g}} := \sum_{t', s'} \c \ket{t', s'}
\end{equation}
where $\ket{t', s'}$ is an eigenstate of the Hamiltonian with energy $t'$ ($s'$ labels the degeneracy). From Eq. (\ref{formPsi}) it follows that the sum in Eq. (\ref{lineardecompPsi}) will be peaked around only a few energy values $t'$. Indeed, with
\begin{equation}
\ket{\phi_1} := a\ket{0} + b\ket{1},
\end{equation}
and
\begin{equation}
\ket{\phi_2} := b \ket{0} - a\ket{1},
\end{equation}
set $\text{Typ} := [ nE_t - \sqrt{n}, nE_t + \sqrt{n} ]$, where $E_t := \left( (n - t)|b|^{2} + t |a|^{2} \right) /n$. Then 
\begin{equation}
\left \Vert  \sum_{t' \notin \text{Typ}_t, s'}\c \ket{t', s'}   \right \Vert = 2^{- \Omega(\sqrt{n})}
\label{eq:tails-av-E}
\end{equation}

Note that since  $E_t := \left( (n - t)|b|^{2} + t |a|^{2} \right) /n$ the
degeneracy of each energy state $\ket{t_k, g_s}$ is at least as large
as the degeneracy of $\ket{\Psi_{ks}}$. 

Now we construct the reference frame.
Let $\ket{w}$ be an energy eigenstate with energy $n(p |b|^{2} + (1 - p)|a|^{2}) - n^{2/3}$. It is needed to pad the dimension of the reference frame, since
although the probability that it happens is vanishingly small, the unitary does
connect states with large energy difference.  After the protocol, we will see that $\ket{w}$ will hardly be changed, and thus
is only used as a catalyst.  We define the reference system as follows
\begin{equation}
\ket{H} := \frac{1}{\sqrt{|H|}} \sum_{h \in H} \ket{h}
\end{equation}
with $\ket{h} := \ket{h'}\otimes \ket{w}$, where$\ket{h'}$ is an energy eigenstate of energy $h'$ and $H := \{ 0, ..., 2 n^{2/3} \}$. 

Consider the energy preserving unitary
\begin{equation}
U := \sum_{h, t, s, t', s'}\c \ket{t', s'}\bra{t_k, s_g} \otimes \ket{h + t - t'}\bra{h}.
\end{equation}
Then in the sequel we prove that
\begin{eqnarray} \label{main}
&& \left \Vert  U \left(\sum_{k \in \text{Typ}_{\rho} ,s} p_k \ket{t_k, s_g}\bra{t_k, s_g} \otimes \ket{H} \bra{H}\right)U^{\cal y}  - \rho^{\otimes n} \otimes \ket{H}\bra{H} \right \Vert_1 \nonumber \\ &\leq&  O(n^{-1/6}).
\end{eqnarray}

We first analyze the action of $U$ in $\ket{s, t} \otimes \ket{H}$:
\begin{eqnarray} 
U \left( \ket{t_k,s_g} \otimes \ket{H}  \right) &=& \sum_{t', s'} \c \ket{t', s'}\otimes \left( \frac{1}{\sqrt{|H|}} \sum_{h \in H} \ket{h + t_k - t'}  \right) \nonumber \\ &=& \ket{\nu_1} + \ket{\nu_2} + \ket{\nu_3}
\end{eqnarray}
where the non-normalized pure states $\ket{\nu_k}$ are given by
\begin{equation}
\ket{\nu_1} := \sum_{t' \in \text{Typ}_t, s'}\c \ket{t', s'} \otimes \ket{H},
\end{equation}
\begin{equation}
\ket{\nu_2} := \sum_{t' \in \text{Typ}_t, s'} \c \ket{t', s'} \otimes \ket{\text{err}_{t'}}
\end{equation}
with
\begin{equation}
\ket{\text{err}_{t'}} := \frac{1}{\sqrt{|H|}} \sum_{h \in H} \ket{h + t - t'}  - \ket{H},
\end{equation}
and
\begin{equation}
\ket{\nu_3} := \sum_{t' \notin \text{Typ}_t, s'} \c \ket{t', s'} \otimes \left( \frac{1}{\sqrt{|H|}} \sum_{h \in H} \ket{h + t - t'}  \right).
\end{equation}
Set $t_k=E_k$ and let us take $s_g=g$.  We can do the latter since as we
mentioned, the degeneracy
of $\ket{t_k,s_g}$ is larger than the degeneracy of  $\ket{\Psi_{k,g}}$.   
Then,
\begin{eqnarray} \label{closetoright}
\Vert U \left( \ket{t_k, s_g} \otimes \ket{H}  \right) - \ket{\Psi_{k, g}} \otimes \ket{H} \Vert &\leq& \Vert \ket{\nu_1} - \ket{\Psi_{k, g}} \otimes \ket{H} \Vert \nonumber \\ &+& \Vert \ket{\nu_2} \Vert + \Vert \ket{\nu_1} \Vert.
\end{eqnarray}

We now show that the three terms in the R.H.S. are small. For $\ket{\nu_2}$ we first note that for $t' \in \text{Typ}_t$
\begin{equation}
\Vert \ket{\text{err}_{t'}} \Vert \leq n^{-1/6}.
\end{equation}
by taking the worst case.
Then
\begin{eqnarray}
\Vert \ket{\nu_2} \Vert ^{2} &=& \sum_{t' \in \text{Typ}_t, s'} |\c|^{2} \Vert \ket{\text{err}_{t'}} \Vert^{2} \nonumber \\ &\leq& \max_{t' \in \text{Typ}_t} \Vert \ket{\text{err}_{t'}} \Vert^{2} \leq n^{-1/3}.
\end{eqnarray}

For $\ket{\nu_3}$, in turn, we have
\begin{eqnarray}
\Vert \ket{\nu_3} \Vert^{2} &\leq& \sum_{t' \notin \text{Typ}, s'} |\c|^{2} \Vert \ket{\text{err}_{t'}} \Vert^{2} \nonumber \\ &\leq&   \sum_{t' \notin \text{Typ}, s'} |\c|^{2} \leq 2^{- \Omega(\sqrt{n})}.
\end{eqnarray}

Finally, for $\ket{\nu_1}$,
\begin{eqnarray}
\Vert \ket{\nu_1} - \ket{\Psi_{k, g}} \otimes \ket{H} \Vert &=& \left \Vert \sum_{t' \in \text{Typ}_t, s'} \c \ket{t', s'}  - \ket{\Psi_{k, g}}  \right \Vert \nonumber \\ &\leq& 2^{- \Omega(\sqrt{n})}.
\end{eqnarray}

From Eq. (\ref{closetoright}) it thus follows that
\begin{equation}
\Vert U \left( \ket{t_k, s_g} \otimes \ket{H}  \right) - \ket{\Psi_{k, g}} \otimes \ket{H} \Vert \leq O(n^{-1/6}).
\end{equation}
Since $\Vert \ket{\psi}\bra{\psi} - \ket{\phi}\bra{\phi} \Vert_1 \leq \sqrt{2}\Vert \ket{\psi} - \ket{\phi} \Vert$ for every two states $\ket{\psi}, \ket{\phi}$, we find
\begin{eqnarray}
&& \left \Vert U \left( \ket{t_k, s_g}\bra{t_k, s_g} \otimes \ket{H}\bra{H}  \right) U^{\cal y} - \ket{\Psi_{k, g}}\bra{\Psi_{k, g}} \otimes \ket{H}\bra{H} \right \Vert_1 \nonumber \\ &\leq& O(n^{-1/6}).
\end{eqnarray}
Eq. (\ref{main}) then follows from the triangle inequality for trace-norm and Eq. (\ref{typpicalPsits}).

\section{Distillation of quasiclassical states in arbitrary dimensions}
\label{sec:distq}

In this section we present the details of the distillation protocol for quasiclassical states for the general case of $d$-dimensional systems. This is presented as an example of how the results can be extended to arbitrary dimensions using arguments very similar to those used in the two-dimensional case. 

The input to the protocol consists of $n$ copies of the initial resource $\rho$ and $\ell$ copies of the Gibbs state $\gamma$ of the same Hamiltonian $H$. Since the states are quasiclassical, the overall state of the input is fully described by a collection of strings $\mathbf{s}\in\{0,\dots,d{-}1\}^{n+\ell}$ listing the energy level occupied by each system, each string weighted by its probability of occurrence. Here $d$ is the total number of energy levels of $H$, and the first $n$ entries of $\mathbf{s}$ correspond to the state of $\rho$ and the remainder correspond to $\gamma$.  If some of the energy levels are degenerate, we simply work in the eigenbasis of $\rho$ to remove the degeneracy in labeling. 
Because permutations within each of the two substrings do not change the overall probability, we can therefore instead work with the collection of occupation frequencies $\mathbf{f}$ of the state, which describe the number of systems in the ground state, first excited state, second excited state, and so on (divided by the total number of systems), again each weighted by an appropriate probability. 

We would now like to define a protocol which creates as many standard resources in the form of work as possible. In the qubit case the standard resource had a very simple form, namely the excited state of the Hamiltonian. Here, however, the setup is more cumbersome, as there are $d-1$ excited states and no guarantees that their energy differences are in any way commensurate (as, e.g. in the case of a harmonic oscillator, where the energy of the state $\ket{2}$ can be transferred to two instances of the state $\ket{1}$). To handle this issue most simply, we imagine that, in addition to the resource and thermal states, we also have a work system at our convenience. The work system is capable of accepting arbitrary amounts of energy, i.e.\ it has energy transitions which precisely correspond to those of $H$, but it cannot accept any entropy. Now the goal of the protocol is to change the occupation of the energy levels so as to transfer as much energy to the work system as possible. 

Let us now restrict attention to a fixed occupation frequency $\mathbf{f}_\rho$ of the resource and a fixed $\mathbf{f}_\gamma$ for the Gibbs state. 
We will later design the protocol so that it works for every such frequency pair which has appreciable probability. Suppose that we now change the occupation numbers by an amount described by the vector $-n\mathbf{x}$ (whose prefactor is chosen for later convenience).  This results in a new occupation vector $\boldsymbol{\nu}$ defined by 
\begin{align}
\label{eq:nudef}
(n+\ell)\boldsymbol{\nu}\equiv n\mathbf{f}_\rho+\ell\mathbf{f}_\gamma-n\mathbf{x}.
\end{align}
For this to be an allowable transformation in our framework, this mapping must satisfy two constraints: energy conservation and unitarity. In contrast to the qubit case, here the input and output dimensions are equal by design. 

Energy conservation is simply enforced by requiring the work system to take up the change in energy of the input systems. Using the vector $\mathbf{H}$ to describe the energy of each energy level, the initial energy is given by $E_{\rm in}= n\mathbf{H}\cdot\mathbf{f}_\rho+\ell\mathbf{H}\cdot\mathbf{f}_\gamma$ while the final energy is $E_{\rm out}=n\mathbf{H}\cdot\mathbf{f}_\rho+\ell\mathbf{H}\cdot\mathbf{f}_\gamma-n\mathbf{H}\cdot\mathbf{x}$. Energy conservation is then the statement that the work extracted is given by $W=E_{\rm in}-E_{\rm out}=n\mathbf{H}\cdot\mathbf{x}$. 

Unitarity is enforced by making sure that the total number of configurations (strings) consistent with each occupation vector is conserved by the process. The total number of possible input strings in this case, $N_{\rm in}$,
is just the product of the multinomial cofficients using the frequency vectors:
\begin{align}
N_{\rm in}=M(n\mathbf{f}_\rho)M(\ell\mathbf{f}_\gamma)=\frac{n!}{(n(\mathbf{f}_\rho)_0)!\cdots(n(\mathbf{f}_\rho)_{d{-}1})!}\frac{\ell!}{(\ell(\mathbf{f}_\gamma)_0)!\cdots(\ell(\mathbf{f}_\gamma)_{d{-}1})!}
\end{align}
The maximum number of strings $N_{\rm out}=M((n+\ell)\boldsymbol{\nu})$ which can be created in the $n+\ell$ systems given the new occupation frequency $\boldsymbol{\nu}$ is the multinomial coefficient of the new occupation frequency vector,
\begin{align}
N_{\rm out}=M((n+\ell)\boldsymbol{\nu})=\frac{(n+\ell)!}{((n+\ell){\nu}_0)!\cdots((n+\ell){\nu}_{d{-}1})!}.
\end{align}
Therefore, a sufficient condition for unitarity is $N_{\rm in}\leq N_{\rm out}$, or $M(n\mathbf{f}_\rho)M(\ell\mathbf{f}_\gamma)\leq M((n+\ell)\boldsymbol{\nu})$.

It can be shown that the multinomial coefficients obey the bounds
\begin{align}
\frac{e}{(ne)^d}\frac{1}{f_1\cdots f_d}2^{nH(\mathbf{f})}\leq \frac{n!}{(nf_1)!\cdots (nf_d)!}\leq \frac{n}{e^d-1}2^{nH(\mathbf{f})},
\end{align}
and therefore $M(n\mathbf{f})\approx 2^{nH(\mathbf{f})\pm O(\log n)}$.
The {unitarity} condition then becomes 
\begin{align}
	\label{eq:unitdef}
nH(\mathbf{f}_\rho)-O(\log n)+\ell H(\mathbf{f}_\gamma)-O(\log \ell)\leq (n+\ell)H(\boldsymbol{\nu})+O(\log (n+\ell))
\end{align}
Defining $\eps=\frac{n}{\ell}$ we may express this as 
\begin{align}
\eps H(\mathbf{f}_\rho) + H(\mathbf{f}_\gamma) \leq (1+\eps)H(\boldsymbol{\nu}) + O(\tfrac{\log \ell}{\ell}).
\end{align}
Using the expression for $\boldsymbol{\nu}$ from \eqref{eq:nudef} and assuming that $\epsilon\ll 1$ gives
\begin{align}
H(\boldsymbol{\nu}) &= H\left(\frac{n\mathbf{f}_\rho+\ell\mathbf{f}_\gamma-n\mathbf{x}}{n+\ell}\right)\\
&=H(\mathbf{f}_\gamma)+\eps\left[\left(\mathbf{f}_\gamma+\mathbf{x}-\mathbf{f}_\rho\right)\cdot\mathbf{1}-H(\mathbf{f}_\gamma)+\left(\mathbf{x}-\mathbf{f}_\rho\right)\cdot\log\mathbf{f}_\gamma\right]+O(\eps^2)\\
&=H(\mathbf{f}_\gamma)-\eps\left(\mathbf{f}_\rho-\mathbf{x}\right)\cdot\log\mathbf{f}_\gamma-\eps H(\mathbf{f}_\gamma)+O(\eps^2).
\end{align}
Here $\mathbf{1}$ is the vector of all ones, and we have made use of the fact that $\mathbf{f}\cdot\mathbf{1}=1$ for any frequency vector $\mathbf{f}$, which also implies $\mathbf{x}\cdot\mathbf{1}=0$. Combining this with \eqref{eq:unitdef} we obtain the relation
\begin{align}
\label{eq:ucondition}
-\mathbf{x}\cdot\log\mathbf{f}_\gamma\leq D(\mathbf{f}_\rho||\mathbf{f}_\gamma)+O(\tfrac{\log \ell}{\ell}).
\end{align}

The next step is to fix the protocol to the worst case among the {likely} frequency vectors $\mathbf{f}_\rho$ and $\mathbf{f}_\gamma$. Their probabilities sharply peaked around the individual distributions $\boldsymbol{\rho}$ and $\boldsymbol{\gamma}$, respectively. Specifically, fixing an error parameter $\delta$, the probability that $||\mathbf{f}_\rho-\boldsymbol{\rho}||_1\geq \delta$ is less than a quantity of order $e^{-n\delta^2}$. 
The variations of likely $\mathbf f_\rho$ from $\boldsymbol \rho$ itself are again $O(\tfrac1{\sqrt{n}})$ as in the argument presented in the main text (there the statement was phrased in terms of the number of $1$'s and not the type class or frequency distribution itself). Thus we may choose $\ell=(Rn^{3/2})$ to ensure that \eqref{eq:nudef} and \eqref{eq:unitdef} hold with $\mathbf f_\rho$ replaced with $\boldsymbol \rho$ and similarly for $\gamma$, at least to terms sublinear in $n$. We conclude that even in the worst, but still probable case, we have
\begin{align}
-\mathbf{x}\cdot\log\boldsymbol{\gamma}\leq D(\boldsymbol{\rho}||\boldsymbol{\gamma})-O(\tfrac{1}{\sqrt{n}}).
\end{align}
Now $\log\gamma =-\beta\mathbf{H}-\mathbf{1}\log Z$, so this condition becomes
\begin{align}
\beta\mathbf{H}\cdot\mathbf{x}\leq D(\boldsymbol{\rho}||\boldsymbol{\gamma})-O(\tfrac{1}{\sqrt{n}}).
\end{align}
This equation gives the minimum amount of extractable work among all the likely frequencies, which is taken to be the target amount for the process. As the extraction is unitary for each frequency, and these correspond to disjoint quantum states, we can thus find a unitary for the entire input capable of generating 
$\tfrac{1}{\beta}\left[D(\boldsymbol{\rho}||\boldsymbol{\gamma})-O(\tfrac{1}{\sqrt{n}})\right]$ units of useful work per input resource state, with probability greater than $1-O(\tfrac{1}{\sqrt{n}})$.

\section{Structure of the exhaust state}

When doing a transformation of $\rho^{\otimes n}$ into $\sigma^{\otimes m}$, at the end of the protocol we actually obtain $\sigma^{\otimes m} \otimes \pi_{k}$ and we trace out $\pi_{k}$, which lives in $k = \Omega(n)$ copies of the system. Although $\pi_n$ is usually far away, in fidelity, from many copies of a Gibbs state, we show that its reductions are very close to a Gibbs state. The main observation is that because $\pi_{n}$ should be useless for extracting more copies of $\sigma$ at a non-zero rate, we must have
\begin{equation}
S(\pi_k || \rho_{\beta}^{\otimes k}) \leq k^{1 - \delta},
\end{equation}
for $\delta > 0$. But by subadditivity of the entropy we have
\begin{eqnarray}
S( \pi_k || \rho_{\beta}^{\otimes k}) &=& - S(\pi_k) - \sum_{l=1}^{k} \tr(\rho_{k, l}\log \rho_{\beta}) \nonumber \\ &\geq& - \sum_{l=1}^{k} S(\pi_{k, l}) - \sum_{l=1}^{k} \tr(\rho_{k, l}\log \rho_{\beta}) \nonumber \\ &=& \sum_{l=1}^{k} S(\pi_{k, l} || \rho_{\beta}),
\end{eqnarray}
where $\pi_{k, l} := \tr_{\backslash l} (\pi_{k})$ is the reduced state of $\pi_k$ that is obtained by partial tracing all the systems except the $l$-th one. Let us assume for simplicity that all the $\pi_{k, l}$ are identical. Then 
\begin{equation}
S(\pi_{k, 1} || \rho_{\beta}) \leq k^{-\delta},
\end{equation}
which by Pinsker's inequality implies
\begin{equation}
\Vert \pi_{k, 1} - \rho_{\beta} \Vert_{1} \leq \Omega(k^{-2 \delta}).
\end{equation}

More generaly, repeating the same argument for larger blocks we get that 
\begin{equation}
\Vert \tr_{L, L+1, ..., k}\left( \pi_{k} \right) - \rho_{\beta}^{\otimes L} \Vert_{1} \leq \Omega(Lk^{-2 \delta}).
\end{equation}

\section{Equivalence and degree of control for thermal operations }

Here, we address two questions. The first is how our paradigm, where we use unitaries $V$ which commute with the total Hamiltonian $H$, relates to other 
approaches.  The second is how much control an experimenter needs over the choice of unitaries $V$.  To answer the first question, consider a common
approach to thermodynamics, which is to manipulate thermodynamical systems using an external apparatus.  In this model, the systems are manipulated using
a time-dependent Hamiltonian, $H(t)$.  Another approach is to add an interaction term $H_{\rm int}$ between various systems we are 
trying to manipulate (e.g. the resource, and the heat bath), and then bring these systems into contact with one another.  Let us now see that these are equivalent
to considering unitaries $V$ which commute with the original Hamiltonian $H$.

First, observe that in the case of a time-dependent Hamiltonian $H(t)$, we can simply include the clock as one of our systems. Letting $\tau$ be the coordinate operator of the clock system and $\Pi_\tau$ such that $[\tau,\Pi_\tau]=-i$, define $H_{\rm indep}=H(\tau)+\Pi_\tau$.  The $\tau$ observable faithfully records the time $t$, as can be seen by solving the Heisenberg
equations of motion to get $\tau(t)=\tau(0)+ t$. Now consider a joint density matrix for the system plus clock of the form $\xi(t)=\rho(t)\otimes \proj t$, where $\ket{t}$ are the eigenstates of $\tau$. The time-independent Hamiltonian $H_{\rm indep}$ acting on the state $\xi(t)$ will generate the equation of motions of $H(t)$ acting on $\rho(t)$, but will also conserve energy. 
To see this, recall that the product rule of derivatives gives
\begin{align}
\frac{d\xi(t)}{dt}=\frac{d\rho(t)}{dt}\otimes\proj{t}+\rho(t)\otimes \frac{d}{dt}\proj{t},
\end{align}
while the Heisenberg equation of motion gives
\begin{align}
\frac{d\xi(t)}{dt}&=i[H_{\rm indep},\xi(t)]\nonumber\\
&=i[H(t),\rho(t)]\otimes\proj{t}+\rho(t)\otimes[\Pi_\tau,\proj{t}]
\end{align}
Comparing the above two equations we have $\dot\rho(t)=i[H(t),\rho(t)]$ as claimed. That a system in a pure state stays in a pure state, can be achieved by having the clock have a large coherent superposition over energy levels, thus the change in it's state can be made arbitrarily small, as explained in Section \ref{sec:form_arb}.
We thus can go from
a picture with a changing Hamiltonian, to one with a fixed one.  The model with time-dependent Hamiltonian is therefore equivalent to the one considered here, with fixed Hamiltonian.

Likewise, in the case where an interaction term is added, we can take the 
total Hamiltonian to be $H_{\rm tot}=H+H_{\rm int}$ and assume that initially, $(H+H_{\rm int})\ket\psi\approx H\ket\psi$ i.e.\ the systems are initially far apart.
They can then evolve unitarily, such that the systems interact, and then move far enough apart that the interaction terms are negligible again.
In such a picture, an eigenstate of the initial Hamiltonian $H$ will evolve into an eigenstate of $H$ with the same energy 
(by conservation of energy, and the fact that the interaction is negligible at initial and final times).  Thus, all that happens here is that
eigenstates of fixed energy evolve to other eigenstates of the same energy, and this can be accomplished by means of a fixed Hamiltonian $H$
and a unitary $V$ which commutes with it.  We thus see that also the picture of adding interaction terms 
is equivalent to having a fixed Hamiltonian $H$, and operations $V$ which commute with it.

Similarly, the application of a unitary during some time period can be made via application of a fixed Hamiltonian.
One can include an internal clock $\tau$ which merely acts as a catalyst and thus
have some fixed Hamiltonian
\begin{align}
H_{\rm tot}=H+H_{\rm int}g(\tau)+\Pi_\tau
\label{eq:unitarytoham}
\end{align}
which effectively implements $e^{-iH_{\rm int}t}$ over some time interval determined by the function $g(\tau)$.
Here $\Pi_\tau$ is conjugate to $\tau$ and one can verify via the Heisenberg equations of motion that $\tau$ depends linearly on $t$.  
Since $[H_{\rm int},H]=0$ one can also verify via the Heisenberg equations of motion for $\Pi_\tau$ that there is no 
backreaction or energy exchange to the clock at late times provided $g(\tau)$ is chosen such that $\int_{t_i}^{t_f} g'(\tau)=0$ and 
$g(\tau)$ and $\tau(t)$ chosen such that $g(\tau)=0$ before $t=t_i$ and after $t=t_f$.

One might be concerned that if we perturb the Hamiltonian slightly, our work extraction will not be robust.  To see this,
let us consider the case where we don't succeed in implementing our unitary exactly, but rather some $H_{\rm int}$ which does not completely 
commute with the original Hamiltonian $[H,H_{\rm int}]=-i\delta$.  Viewed internally, will see that this is equivalent to allowing a violation of conservation of energy by amount  $\delta$ -- something which is interesting to study in its own right.  Now the equations of motion for $\tau$ are unchanged, and thus 
the unitary $e^{-iH_{\rm int}t}$ is still implemented.  However there is some backreaction on the clock.  Solving the Heisenberg equations of motion for 
$\Pi$, we find that there is a small momentum kick to the clock 
\begin{align}
\Pi(t_f)-\Pi(t_i) &=\int^{t_f}_{t_i} g'(t) H_{\rm int}(t)\nonumber \\
&= H_{\rm int}(t_f)- H_{\rm int}(t_i)
\end{align}
where we have taken $g(\tau)$ to be $1$ between $t_i$ and $t_f$ and $0$ everywhere else.  At each cycle some amount of energy gets stored in the clock, depending
on the initial and final states of the system.  If we allow these to fluctuate, then the transfer is some $\delta$ and the clock  undergoes a random walk.

We thus find that if we run the extraction process as a cycle, where we repeat the process over several cycles, then at each cycle we still exactly
implement  $e^{-iH_{\rm int}t}$, it's just that it now has some tiny non-commuting part with the original Hamiltonian, resulting in some energetic backreaction to the clock, and thus some entropy being stored in 
the clock. However, the extracted work grows linearly, and $\delta$ can be made arbitrarily small.  After $n$ cycles, 
the moment of the clock has undergone a random walk, of order $\sqrt{n}$, an amount which is negligible compared to the extractable work in the
case of many cycles.

Let us now turn to the second question.   It might appear that an experimenter who wished to implement our protocols would need
to very carefully manipulate all the many degrees of freedom of the $n$ systems and the heat bath.  However, this is not the case, as we will now demonstrate explicitly using the example of work distillation.  There, we were mapping eigenstates which had a type $lq$ on the heatbath $\gamma$, and $pn$ on the resource  $\rho$
to microstates which had type $rk$ on the garbage $\sigma$, and $m$ $1$'s on the work system.  However, although for any implementation
of the protocol, we need a particular mapping of strings (i.e.\ microstates) of these initial types, to strings of the final type, 
{\bf any mapping} will do.  The only important thing which is required is just that the unitary operation 
map the initial types to the final types.  Thus an experimenter who wishes to
implement the protocol, does not need fine-grained control over the mapping of microstates within one type to microstates within another type.  
She only needs to know that the unitary maps one type into another.  In other words, there are an exponentially large number of 
possible implementations of our protocols each of which
map particular strings within the initial types to particular strings in the final types.  However, it doesn't matter which implementation is chosen,
and the experimenter thus does not need the fine degree of control that is required to achieve a particular implementation.

We can think of the type as being like a macroscopic variable such as the total magnetisation of a composite system, or its total energy (indeed it is the latter). 
In the distillation protocol, we map the macroscopic variables of energy on two large systems ($\gamma^{\otimes l} $ and $\rho^{\otimes n}$) to the macroscopic
variable of energy on the final system.  Any unitary which accomplishes this will successfully implement our protocol.  Thus, the experimenter only
needs control over the macroscopic variables, not the microscopic ones.

Equivalence of these paradigms is discussed in more detail, and in the case of finite systems, in \cite{HOcrude}.

\end{document}